\documentclass[showpacs,prd,twocolumn,floatfix,
nofootinbib,superscriptaddress]{revtex4}
\usepackage{graphicx}

\newcommand{\be}{\begin{equation}}
\newcommand{\ee}{\end{equation}}
\newcommand{\nl}{\nonumber \\}

\begin{document}


\title{$D \rightarrow \pi, l \nu$  Semileptonic Decays, $|V_{cd}|$
 and 2$^{nd}$ Row Unitarity from Lattice QCD}

\author{Heechang Na}
\affiliation{Department of Physics,
The Ohio State University, Columbus, OH 43210, USA}
\author{Christine T.\ H.\ Davies}
\affiliation{SUPA, School of Physics \& Astronomy, 
University of Glasgow, Glasgow, G12 8QQ, UK}
\author{Eduardo Follana}
\affiliation{Departamento de Fisica Teorica, 
Universidad de Zaragoza, E-50009 Zaragoza, Spain} 
\author{Jonna Koponen}
\affiliation{SUPA, School of Physics \& Astronomy,
University of Glasgow, Glasgow, G12 8QQ, UK}
\author{{G.Peter} Lepage}
\affiliation{Laboratory of Elementary Particle Physics,
Cornell University, Ithaca, NY 14853, USA}
\author{Junko Shigemitsu}
\affiliation{Department of Physics,
The Ohio State University, Columbus, OH 43210, USA}

\collaboration{HPQCD Collaboration}
\noaffiliation


\begin{abstract}
We present a new calculation of the $D \rightarrow \pi, l \nu$ semileptonic 
form factor $f^{D \rightarrow \pi}_+(q^2)$ at $q^2 = 0$ based on HISQ charm and light 
valence quarks on MILC $N_f = 2 +1$ lattices. Using methods developed recently 
for HPQCD's study of $D \rightarrow K, l \nu$ decays, we find 
$f^{D \rightarrow \pi}_+(0) = 0.666(29)$.  This signifies a better than factor of two  
improvement in errors for this quantity compared to previous calculations. 
Combining the new result with CLEO-c branching fraction data, we extract 
the CKM matrix element $|V_{cd}| = 0.225(6)_{exp.}(10)_{lat.}$, where the first error comes from 
experiment and the second from theory. With a total error of $\sim5.3$\%  the accuracy 
of direct 
determination of $|V_{cd}|$ from $D$ semileptonic decays has become comparable to 
(and in good agreement with) that from neutrino scattering.  
 We also check for second row 
unitarity using this new $|V_{cd}|$,  HPQCD's earlier $|V_{cs}|$ and 
$|V_{cb}|$ from the Fermilab Lattice \& MILC collaborations. We find 
$|V_{cd}|^2 + |V_{cs}|^2 + |V_{cb}|^2 = 0.976(50)$, improving on 
the current PDG2010 value.
\end{abstract}


\maketitle


\section{Introduction}
The Cabibbo-Kobayashi-Maskawa (CKM) matrix provides particle physicists with 
a wealth of opportunities to carry out precision tests of 
the Standard Model (SM) and look for New  Physics.  On the 
one hand  each matrix element can be determined in several  ways, 
employing different experimental and theory inputs, and the results 
compared with each other.  Three generation unitarity can also be examined 
to see how well $\hat{V}_{CKM} \times \hat{V}_{CKM}^\dagger = \hat{I}$ 
is satisfied.  This leads to tests such as 1$^{st}$, 2$^{nd}$ or 3$^{rd}$ 
row/column unitarity.  It also gives rise to the important ``Unitarity Triangle'' 
relation $V_{ud} * V^*_{ub} + V_{cd}*V^*_{cb} + V_{td}*V^*_{tb} = 0$.  Consistency 
checks of the sides and angles of the Unitarity Triangle (UT) have been the focus 
of much of the experimental and theoretical effort in Flavor Physics in 
recent years.
Lattice QCD is playing an increasingly important role in 
CKM physics \cite{reviews}.
 For instance, lattice calculations of the Kaon 
semileptonic form factor $f_+^{K \rightarrow \pi}(0)$ \cite{boyle,lubicz}
 and the decay constant $f_K$ (or $f_K/f_\pi$) \cite{milc,follana,bmw,rbc,etm} 
have contributed to precision 
determinations of $|V_{us}|$ and 1$^{st}$ row unitarity tests \cite{kaonrev}.

The HPQCD collaboration recently published a new lattice calculation of the $D 
\rightarrow K, l \nu$ semileptonic decay form factor $f^{D \rightarrow K}
_+(q^2)$ at $q^2 = 0$ \cite{dtok}
which significantly reduced the error on this quantity compared to 
previous theory results. This led to a very precise determination of 
the CKM matrix element $|V_{cs}|$.  Features in the HPQCD work that made this 
 improvement possible include the use of a relativistic quark action, 
the ``Highly Improved Staggered Quark'' (HISQ) action \cite{hisq}, to simulate both light 
and charm quarks, better data analysis tools and a new method for 
carrying out chiral/continuum extrapolations. 
Capitalizing on these developments, we turn here in this article to 
$D \rightarrow \pi, l \nu$ semileptonic decays. We focus on extracting the 
CKM matrix element $|V_{cd}|$ by combining theory results for 
$f_+^{D \rightarrow \pi}(0)$ with experimental input from CLEO-c \cite{cleo}. The first 
unquenched lattice studies of $D$ semileptonic decays 
 were carried out several years ago by the Fermilab Lattice \& 
MILC collaborations \cite{fermi2005}. In that work lattice gauge theory was able to predict 
the shape of $f_+(q^2)$ prior to subsequent confirmation by 
experiment.  The theory errors for $f_+^{D \rightarrow \pi}(0)$ 
in \cite{fermi2005} were $\sim$10\%, and this has remained the dominant 
error in determinations of $|V_{cd}|$ from $D$ semileptonic decays.  More accurate 
determinations have come from neutrino scattering experiments so that 
the current PDG2010
 \cite{pdg2010} quotes $|V_{cd}|$ from neutrino charm production with 
 an error of about $\sim$5\%.  With the new lattice calculations 
described in this article, the accuracy 
of $|V_{cd}|$ from $D$ semileptonic decays is approaching that from 
neutrino scattering and this provides an important consistency check. We find,
\be
\label{vcd}
|V_{cd}| = 0.225(6)_{exp.}(10)_{lat.},
\ee
where the first error is from experiment \cite{cleo} and the second is the 
theory error from the Lattice QCD calculation presented here. Eq.(\ref{vcd}) 
is in excellent agreement with the PDG value based on neutrino scattering 
of $|V_{cd}| = 0.230(11)$. 

In the rest of this article we describe how the result of 
eq.(\ref{vcd}) was obtained. 
 We work with HISQ valence charm and light quarks 
on the MILC AsqTad $N_f = 2 + 1$ coarse ($a \sim 0.12$fm) and fine 
($a \sim 0.09$fm) lattices. 
 Table I lists the five MILC \cite{milc1} ensembles employed 
in this work and some simulation parameters.  Compared to \cite{dtok} 
we have doubled the statistics on ensembles C1, C2 and F1. 
The valence charm and light bare quark masses are the same as in \cite{dtok} with the former 
tuned to the $\eta_c$ mass and the latter chosen such that the ratio of 
$m_{light}$ to the physical strange quark mass is approximately the same 
for valence and sea quarks.   
In the next section we summarize the formulas for hadronic matrix elements 
necessary to extract $f^{D \rightarrow \pi}_+(0)$ and explain how they are related 
to three- and two-point correlators evaluated numerically on the lattice. 
 These relations are the same as those described in reference \cite{dtok} 
so we will be brief. In section III we describe our data analysis and 
fitting procedures.  We employ Bayesian fitting methods 
and carry out multi-exponential 
fits to several three-point and two-point correlators at the same time.  
Section IV discusses chiral and continuum extrapolations of lattice results 
to the physical limit. We apply the modified z-expansion method developed for 
$D \rightarrow K$ semileptonic form factors in \cite{dtok}.
Section V presents our results for $f^{D \rightarrow \pi}_+(0)$ and 
$|V_{cd}|$ and comparisons with other determinations of these quantities.  
 Section VI gives a brief summary and we also include a 
$2^{nd}$ row unitarity test with all theory inputs coming from Lattice QCD.

\begin{table}
\caption{ The MILC $N_f = 2+1$ ensembles
 used in the $D \rightarrow \pi$ semileptonic project. 
The fourth column gives the valence HISQ light and charm 
quark masses in lattice units.  $N_{conf}$ is the number 
of configurations and $N_{tsrc}$ the number of time sources 
used for each configuration.
}
\begin{center}
\begin{tabular}{|c|c|c|c|c|c|c|}
\hline
Set &$r_1/a$ & $m_l(sea)/m_s(sea)$  & $a m_{valence}$ &  $N_{conf}$&
 $N_{tsrc}$ & $L^3 \times N_t$ \\
\hline
\hline
C1  & 2.647 & 0.005/0.050 & 0.0070   & 1200  &  2 & $24^3 \times 64$ \\
    &&&                    0.6207   & 1200  &   2 & \\
\hline
C2  & 2.618 & 0.010/0.050 & 0.0123   & 1200 &  2 & $20^3 \times 64$ \\
    &&&        0.6300   & 1200 &  2 & \\
\hline
C3  & 2.644 & 0.020/0.050 & 0.0246  &  600 &  2 & $20^3 \times 64$ \\
   &&&         0.6235  &  600 &   2 & \\
\hline
\hline
F1  & 3.699 & 0.0062/0.031 & 0.00674 & 1200 &  4  & $28^3 \times 96$ \\
    &&&        0.4130  & 1200 &   4  & \\
\hline
F2  & 3.712 & 0.0124/0.031 & 0.0135 &  600 &  4 & $28^3 \times 96$ \\
    &&&        0.4120  &  600 &  4 & \\
\hline
\end{tabular}
\end{center}
\end{table}

\section{ Relevant Matrix Elements}
The most efficient way to calculate 
$f_+(q^2)$  at $q^2 =0$ is to focus on the scalar form factor 
$f_0(q^2)$ and use the kinematic identity $f_+(0) = f_0(0)$.  The scalar 
form factor can be determined from the matrix element of the scalar 
current $S = \overline{\Psi}_q \, \Psi_c$ between the $D$ meson and 
pion states.
\be
\label{f0}
f^{D \rightarrow \pi}_0(q^2) =
 \frac{(m_{0c} - m_{0l}) \langle \pi | S | D \rangle}{M_D^2 - M_\pi^2}.
\ee
The combination in the numerator of eq.(\ref{f0}) does not get 
renormalized. The use of absolutely normalized currents is one of 
the reasons why we are able to significantly reduce  errors in our 
$D$ semileptonic scalar form factor calculations, both here and in 
\cite{dtok}.

Our goal is to determine the hadronic matrix element $\langle \pi | S | D \rangle$
in eq.(\ref{f0}) via numerical simulations.  The starting point is the
three-point correlator,
\begin{eqnarray}
\label{thrpnt}
& & C^{3pnt}(t_0,t,T,\vec{p}_\pi) =
\frac{1}{L^3} \sum_{\vec{x}}
\sum_{\vec{y}} \sum_{\vec{z}} e^{i\vec{p}_\pi \cdot (\vec{z} - \vec{x})} \nl
&& \qquad \langle \Phi_\pi(\vec{x},t_0) \,\tilde{S}(\vec{z},t) \,
\Phi^\dagger_D(\vec{y},t_0-T)
 \rangle.
\end{eqnarray}
In eq.(\ref{thrpnt}) the interpolating operator $\Phi^\dagger_D$ creates a $D$ meson 
at time slice $t_0 - T$.  At time $t$ ($t_0 \geq t \geq t_0 - T$) the scalar 
current $S$ converts the $c$ quark inside the $D$ into a light quark and also inserts 
momentum $\vec{p}_\pi$.  
 The resulting pion 
then propagates to time slice $t_0$ where it is annihilated by $\Phi_\pi$. 
In addition to the three-point correlator one needs the pion and $D$ meson
two-point correlators,
\be
\label{twopntd}
C^{2pnt}_D(t,t_0) =\frac{1}{L^3} \sum_{\vec{x}} \sum_{\vec{y}}
\langle \Phi_D(\vec{y},t) \Phi^\dagger_D(\vec{x},t_0) \rangle,
\ee
and
\begin{eqnarray}
 & &C^{2pnt}_\pi(t,t_0; \vec{p}_\pi) = \nl
\label{twopntpi}
& & \quad \frac{1}{L^3} \sum_{\vec{x}} \sum_{\vec{y}}
e^{i \vec{p}_\pi \cdot (\vec{x} - \vec{y})}\langle \Phi_\pi(\vec{y},t)
 \Phi^\dagger_\pi(\vec{x},t_0) \rangle.
\end{eqnarray}
Details on how the above three- and two-point correlators can be expressed in terms 
of single component staggered quark propagators are given in section IV of reference 
\cite{dtok} and will not be repeated here.  There we also show how the
sums $\sum_{\vec{x}}$ 
in eqns.(\ref{thrpnt}), (\ref{twopntd}) and (\ref{twopntpi}) can be 
carried out using U(1) random wall sources.  

The meson creation operators $\Phi^\dagger_D$ and $\Phi_\pi^\dagger$ create not only
the ground state $D$ and pion we are interested in but also an arbitrary number 
of excited states with the same quantum numbers.  Hence the $t$ dependence of 
the two- and three-point correlators is complicated especially for staggered quarks. 
For two-point correlators it is given by,
\begin{eqnarray}
C^{2pnt}_D(t) &=& \sum_{j=0}^{N_D-1} b^D_j (e^{-E^D_j t} + e^{-E^D_j ( N_t - t)}) \nl
\label{twopntfit}
&+& \sum_{k=0}^{N_D^\prime - 1} d^D_k (-1)^t
( e^{-E^{\prime D}_k t} + e^{-E_k^{\prime D} (N_t - t)}),\nl
\end{eqnarray}
and similarly for $C^{2pnt}_\pi(t)$, except that there is no opposite parity terms for zero momentum. For three-point correlators one has 
\begin{eqnarray}
\label{thrpntfit}
& &C^{3pnt} (t,T)
 = \sum_{j=0}^{N_\pi-1} \sum_{k=0}^{N_D-1} A_{jk} e^{-E_j^\pi t} e^{ -E_k^D (T-t)} \nl
& & \quad + \sum_{j=0}^{N_\pi-1} \sum_{k=0}^{N_D^\prime - 1} B_{jk}
 e^{-E_j^\pi t} e^{ -E_k^{\prime D} (T-t)} (-1)^{(T-t)} \nl
& & \quad + \sum_{j=0}^{N_\pi^\prime - 1} \sum_{k=0}^{N_D-1} C_{jk}
 e^{-E_j^{\prime \pi} t} e^{ -E_k^D (T-t)} (-1)^t \nl
& & \quad + \sum_{j=0}^{N_\pi^\prime - 1} \sum_{k=0}^{N_D^\prime - 1} D_{jk} e^{-E_j^{\prime \pi} t}
 e^{ -E_k^{\prime D} (T-t)} (-1)^t (-1)^{(T-t)}. \nl
\end{eqnarray}
We are interested in the ground state contributions with amplitudes,
\be
\label{b0d}
b^D_0 \equiv \frac{|\langle \Phi_D | D \rangle|^2}{2 M_D a^3},
\ee
\be
\label{b0pi}
b^\pi_0 \equiv \frac{|\langle \Phi_\pi | \pi \rangle|^2}{2 E_\pi a^3},
\ee
and
\be
\label{a00}
A_{00} \equiv \frac{\langle \Phi_\pi|\pi\rangle \, \langle \pi|S|D\rangle \, \langle D|\Phi_D \rangle}
{(2 E_\pi a^3) \, (2 M_D a^3)} \, a^3.
\ee
So the hadronic matrix element $\langle \pi | S | D \rangle$ that enters 
into the formula for $f^{D \rightarrow \pi}_0(0)$ in (\ref{f0}) is given by,
\be
\label{pisd}
\langle \pi | S | D \rangle = 2 \sqrt{ M_D E_\pi} \; \frac{A_{00}}{\sqrt{b_0^\pi b_0^D}}.
\ee
We have accumulated simulation data for zero momentum $D$ correlators and for 
pion correlators with momenta 
$\frac{2 \pi}{L}(0,0,0)$,
$\frac{2 \pi}{L}(1,0,0)$,
$\frac{2 \pi}{L}(1,1,0)$ and 
$\frac{2 \pi}{L}(1,1,1)$.
The corresponding three-point correlators  were calculated for $T=15, 16$
on the coarse and for $T = 19,20,23$ on the fine ensembles.
In the next section we describe how the combination on the right-hand-side 
of (\ref{pisd}) is obtained from three- and  two-point correlators.

\section{ Fits and Data Analysis}

\begin{figure}
\includegraphics*[height=9.5cm,angle=270]{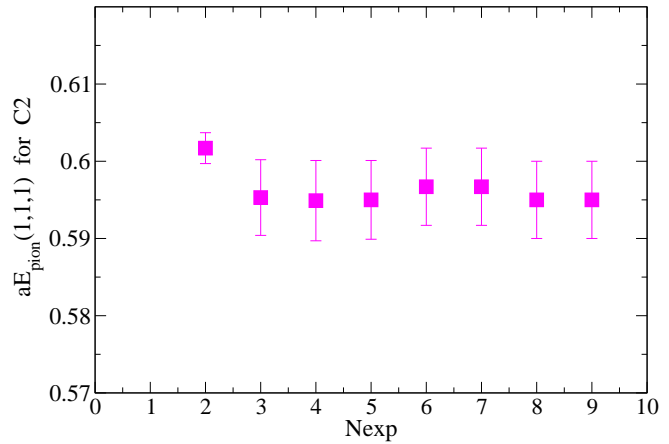}
\caption{
Ground state pion energy in lattice units for momentum $\frac{2 \pi}{L}(1,1,1)$ versus 
the number of exponentials in the fit.
}
\end{figure}

\begin{figure}
\includegraphics*[height=9.5cm,angle=270]{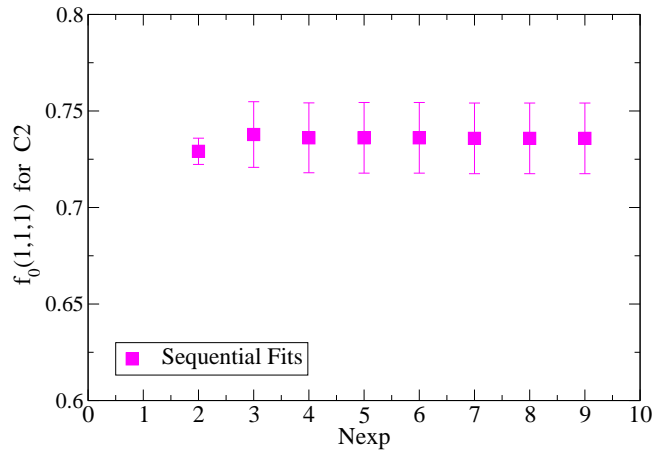}
\caption{
The form factor $f_0(q^2)$ for pion momentum $\frac{2 \pi}{L}(1,1,1)$ from 
simultaneous fits to two- and three-point correlators versus 
$N_{exp} = N_D = N^\prime_D = N_\pi = N^\prime_\pi$. The ``sequential 
fitting'' method  was employed to go from one $N_{exp}$ value to the next. 
In this method the fit results from the $N_{exp}$ exponential fit is
 inserted as  initial conditions for the subsequent [$N_{exp}$ + 1] exponential fit.
}
\end{figure}

Extracting energies and amplitudes from numerical data on two- and three-point correlators
is one of the more challenging but at the same time
 very important aspect of lattice calculations. 
For the past decade the HPQCD collaboration has been employing fitting methods 
based on Bayesian statistics and involving multi-exponential fits \cite{bayse}. 
For instance, in order to obtain the ground state 
energy and amplitude from a two-point correlator, we drop the first 1 $\sim$ 4 time 
slices and then fit to the form of eq.(\ref{twopntfit}) for several values of 
$N \equiv N_D$ (or $N_\pi$) and $N^\prime$.  One continues 
to increase the number of exponentials until the fit results for $E_0$ and $b_0$ 
including their errors and the chisquared per degree of freedom of the fit 
have stabilized.  Fig.1 shows an example of $aE_\pi$ versus $N_{exp} = N_\pi$ 
for momentum $\frac{2 \pi}{L}(1,1,1)$ with 
$N^\prime_\pi = N_\pi$.  One sees that fits have stabilized 
after $N_{exp} = 3$.  It should be noted that as one increases the number of exponentials 
and with it the number of fit parameters, the number of data points is 
growing as well. Each new fit parameter adds another prior term, i.e. additional data,  
to the fit function and the number of data points minus the number of
fit parameters remains constant \cite{bayse}.

The combination $\frac{A_{00}}{\sqrt{b_0^\pi b_0^D}}$ is obtained from 
simultaneous fits to $C^{2pnt}_D(t)$, $C^{2pnt}_\pi(t)$ and 
$C^{3pnt}(t,T)$ for 2 (or 3) $T$ values.  In order to be able to include 
a large number of exponentials in these complicated fits we proceed as 
follows.  We set $N_\pi = N^\prime_\pi = N_D = N^\prime_D \equiv N_{exp}$ 
and start out with a small value, $N_{exp} = 2$ or $N_{exp}=3$.  The fit 
results are then inserted as initial conditions for the subsequent 
$N_{exp} + 1$ exponential fit. This procedure is repeated until 
one has completed multi-exponential fits with $N_{exp}$ much larger than 
2.
Fig. 2 shows results for $f_0(q^2)$ on ensemble C2 at pion momentum 
$\frac{2 \pi}{L}(1,1,1)$ versus $N_{exp}$ using this ``sequential 
fitting'' procedure.  One sees that, similar to in Fig.1, fit results 
have stabilized for $N_{exp} > 3$ \cite{fnote1}. 

In ongoing work we are investigating further methods to deal with 
complicated fits with large number of parameters, in particular fits to collections of 
sums of exponentials \cite{marginal}.
 For the calculations of this article, however, we have found that 
the ``sequential fitting'' method described above works well for all our 
data \cite{fnote2}. 
 We are even able to fit data on a given ensemble for all four 
pion momenta simultaneously and this allows us to obtain correlations 
between form factor results at different $q^2$.  
These simultaneous fit results for $f_0(\vec{p}_\pi)$
 are given in Table II for several pion 
momenta $\vec{p}_\pi$ (the latter in units of $\frac{2 \pi}{L})$.

\section{ Chiral and Continuum Extrapolation }

The next step is to extrapolate the data of Table II to the chiral/continuum 
limit. We do so using the ``modified z-expansion fit'' developed in \cite{dtok}.
The scalar form factor is parameterized as

\begin{eqnarray}
\label{f0fit}
f_0(q^2) & = & \frac{1}{P(q^2) \Phi_0} 
\left ( a_0 D_0 + a_1 D_1 z + a_2 D_2 z^2 \right ) \nonumber \\
 & &  \times \left ( 1 + b_1 (a E_\pi)^2 + b_2 (aE_\pi)^4 \right ),
\end{eqnarray}
with
\begin{eqnarray}
\label{di}
D_i &=& 1 + c_1^i x_l + c_2^i x_l log(x_l) + d_i (am_c)^2 + e_i(am_c)^4 \nonumber \\
  && + f_i \left ( \frac{1}{2} \delta M_\pi^2 + \delta M_K^2 \right ) ,
\end{eqnarray}
\be
x_l = \frac{M_\pi^2}{(4 \pi f_\pi)^2},
\ee
\be
\delta M_\pi^2 = \frac{1}{(4 \pi f_\pi)^2} \left ( (M_\pi^{sea})^2 
- (M_\pi^{valence})^2 \right ),
\ee
\be
\delta M_K^2 = \frac{1}{(4 \pi f_\pi)^2} \left ( (M_K^{sea})^2 
- (M_K^{valence})^2 \right ).
\ee
The kinematic variable $z$ is defined as \cite{bgl,arnesen,bhill},
\be
z(q^2,t_0) = \frac{\sqrt{t_+ - q^2} - \sqrt{t_+ - t_0}}
{\sqrt{t_+ - q^2} + \sqrt{t_+ - t_0}},
\ee
with $t_0$ a free parameter (which we set equal to $1.95$GeV$^2$ ) and $t_\pm = (M_D \pm M_\pi)^2$.  
  We take $\Phi_0$ from \cite{arnesen} and 
set $P(q^2) = 1$, where the latter relation reflects the absence of subthreshold poles in the scalar channel.

\begin{table}
\caption{Results for $f_0(\vec{p}_\pi)$ for each ensemble
}
\begin{center}
\begin{tabular}{|c|c|c|c|c|}
\hline
Set     &  $f_0(0,0,0)$  & $f_0(1,0,0)$ &  $f_0(1,1,0)$ & $f_0(1,1,1)$  \\
\hline
C1   & 1.1557(74) & 0.9155(93)  & 0.8119(68)  & 0.7384(209) \\
C2   & 1.1014(38) & 0.8700(59)  & 0.7801(43)  & 0.7315(87)  \\
C3   & 1.0398(28) & 0.8787(36)  & 0.7922(31)  & 0.7326(66)  \\
F1   & 1.1053(29) & 0.8652(52)  & 0.7586(53)  & 0.7112(108)  \\
F2   & 1.0443(28) & 0.8613(31)  & 0.7645(53)  & 0.7097(70)  \\
\hline
\end{tabular}
\end{center}
\end{table}

\begin{figure}
\includegraphics*[height=9.5cm,angle=270]{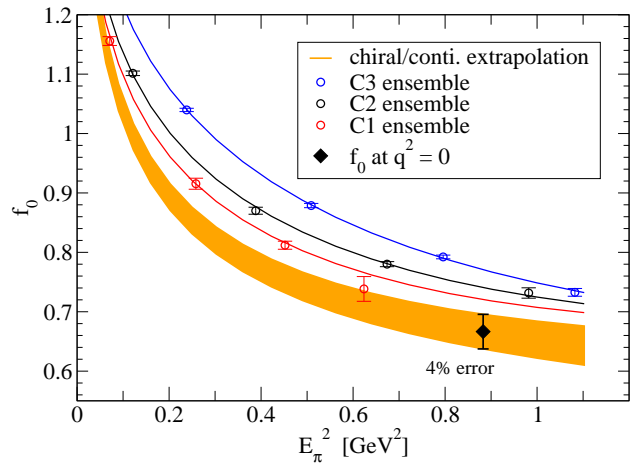}
\caption{
Chiral/continuum extrapolation of $f^{D \rightarrow \pi}_0$ 
versus $E_\pi^2$.  The data points are from the coarse 
ensembles (C1, C2 and C3).  The three individual 
curves and the extrapolated band are from a fit to all five ensembles.
}
\end{figure}

\begin{figure}
\includegraphics*[height=9.5cm,angle=270]{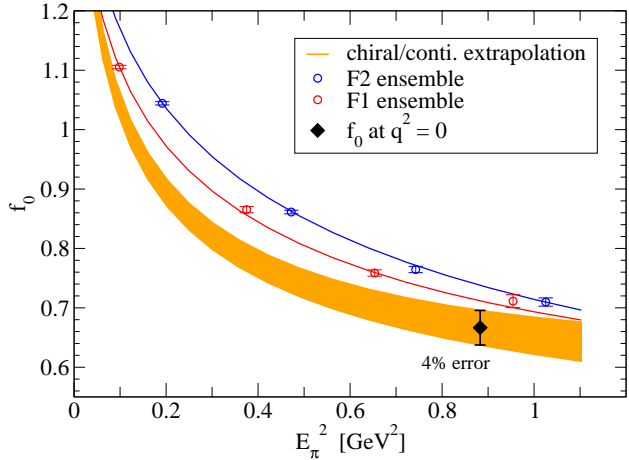}
\caption{
Same as for Fig.3 showing, however, the fine data points.
}
\end{figure}

\begin{figure}
\includegraphics*[height=9.5cm,angle=270]{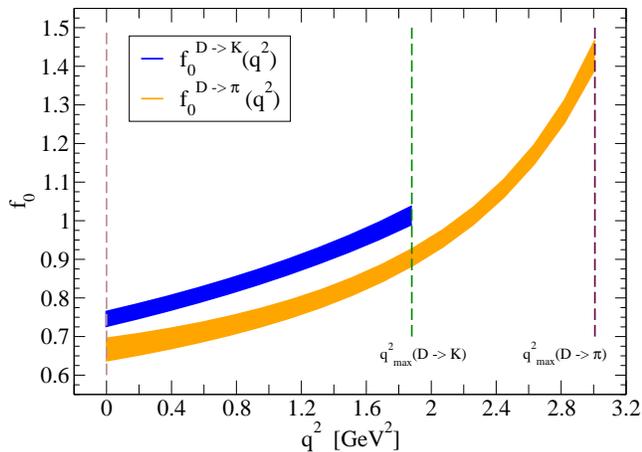}
\caption{
$f_0^{D \rightarrow K}(q^2)$ and $f_0^{D \rightarrow \pi}(q^2) $
versus $q^2$ in the physical limit.
}
\end{figure}

We show results of fits to the form of eq.(\ref{f0fit}) in Figs.3 \& 4. 
We plot separately the coarse and fine data points in order to be able to better 
distinguish individual curves. However, the fit was done simultaneously to 
all the data in Table II, coarse and fine. The $\chi^2/dof = 0.85$ for 
this fit.  
In Fig.5 we show both 
$f_0^{D \rightarrow K}(q^2)$ (results from \cite{dtok})
 and $f_0^{D \rightarrow \pi}(q^2)$ 
versus $q^2$ in the physical region $0 \leq q^2 \leq q^2_{max} = 
(M_D - M_{\pi(K)})^2$.
Note that we consider the correlations between the momenta in the fits which 
we did not consider in \cite{dtok}. However, we find that including or excluding correlations in our chiral/continuum extrapolations has minimal effect on $f_0^{D \rightarrow \pi}(0)$ at the physical point, namely a $\sim0.04 \sigma$ shift in the central value and a $\sim0.02 \sigma$ change in the error. 

The motivation for the ``modified z-expansion fit'' is explained 
in more detail in \cite{dtok}.  Form factors at $q^2 = 0$ are needed to extract 
 the CKM matrix elements $|V_{cd}|$ or $|V_{cs}|$. 
The pion energy approaches 1GeV in this kinematic region and so chiral 
perturbation theory might cease to be valid. 
The z-expansion, on the other hand, is applicable throughout the 
physical kinematic region, and our new ``modified z-expansion'' allows 
for the expansion coefficients to be mass and lattice spacing dependent.
We have checked that
 fits to eq.(\ref{f0fit}) are stable with respect to adding further terms 
such as $x_l^2$, $(am_c)^6$, $(aE_\pi)^6$ or keeping just the $c^i_1$ and 
$c^i_2$ terms in eq.(\ref{di}) (the $D_i$'s). Such changes in the fit ansatz 
led to minimal changes in both the central value and the error for $f_0(0)$ in 
the physical limit.  We have also verified that traditional ChPT extrapolations 
(see Appendix C \& D of \cite{dtok} and references therein for 
relevant ChPT formulas) lead to $f_+(0)$ in the physical limit 
consistent with the z-expansion result and  
 with comparable errors, however with worse $\chi^2/dof$.

\section{Results for $f^{D \rightarrow \pi}_+(0)$ and 
$|V_{cd}|$ in the Physical Limit}

\begin{table}
\caption{ Error Budget for $f^{D \rightarrow \pi}_+(0)$
}
\begin{center}
\begin{tabular}{|c|c|}
\hline
Type  &  Error (\%) \\
\hline
\hline
Statistical   &  3.1 \\
Scale ($r_1$ and $r_1/a$)  &  0.7 \\
Expansion coeff. $a_i$ &  0.3 \\
$E_p$   & 0.6 \\
Light quark dependence  &  1.9 \\
Sea  quark dependence  &  0.6 \\
$am_c$ corrections & 2.0 \\
$aE_\pi$ corrections & 1.0 \\
Finite volume & 0.04 \\
Charm mass tuning  &  0.05 \\
\hline
Total  & 4.4\% \\
\hline
\end{tabular}
\end{center}
\end{table}

Our final result for the $D \rightarrow \pi$ form factor at $q^2 = 0$
averaged over $D^0 \rightarrow \pi^-$ and $D^+ \rightarrow \pi^0$, is,
\be
\label{result}
f_+^{D \rightarrow \pi}(0) = 0.666(20)_{stat.}(21)_{sys.},
\ee
where the first error is statistical and the second systematic.
 Fig.6 plots our new result together with
other theory calculations \cite{fermi2005,sumrule,etmcf0}
and experimental determinations
\cite{cleo,belle}
(the latter use CKM unitarity values for $|V_{cd}|$ to extract $f^{D \rightarrow \pi}
_+(0)$).

 The total error in our $f^{D \rightarrow \pi}_+(0)$
is $4.4$\%, signifying a better than factor of 2 improvement over previous
lattice determinations.
The full error budget is given in Table III.  The largest error is statistical
followed by $am_c$ and light quark mass dependence errors. All but the last
two entries in Table III were obtained using the methods described in
reference \cite{alpha}
and Appendix B of \cite{dtok}. For instance, the ``light quark dependence'' errors come from the $c_1^i$ and $c_2^i$ terms in the fit ansatz eq.(13), the ``$am_c$ corrections'' from the $d_i$ and $e_i$ terms etc.

Finite volume errors were estimated by calculating a pion tadpole integral
both at finite and at infinite volume.  The charm mass tuning error is taken
to be the same as for $D \rightarrow K, l \nu$  for which calculations
at two values of $am_c$ were carried out explicitely to estimate this error.
 Effects from electromagnetism/isospin
breaking and charm sea are expected to give negligible contribution to
the error budget compared to our other errors (see \cite{dtok}).

Finally one can combine our result for $f^{D \rightarrow \pi}_+(0)$ with
CLEO-c's measurement of $|V_{cd}| \times f^{D \rightarrow \pi}_+(0)$ \cite{cleo}
to extract a precision value for  $|V_{cd}|$  from D semileptonic
decays. This leads to the result quoted already in eq.(\ref{vcd}), which
is shown in Fig.7 together with values from PDG2010.

\section{Summary}

\begin{figure}
\includegraphics*[height=9.5cm,angle=270]{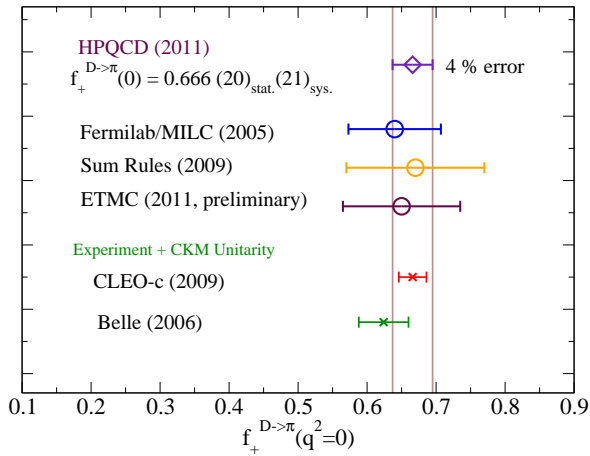}
\caption{
The $D \rightarrow \pi$ form factor $f^{D \rightarrow \pi}_+(0)$ 
from this work and comparisons with other determinations 
\cite{fermi2005,sumrule,etmcf0,cleo,belle}.
}
\end{figure}

\begin{figure}
\includegraphics*[height=9.5cm,angle=270]{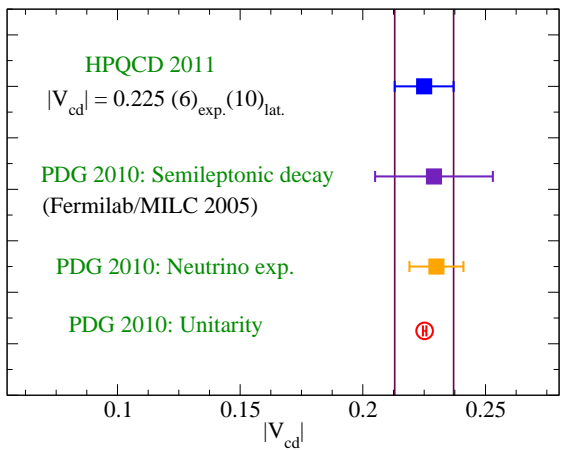}
\caption{
Comparison of $|V_{cd}|$ from this work with values in PDG2010 \cite{pdg2010}.}
\end{figure}

In this article we have  presented a new calculation of the $D \rightarrow \pi, l \nu$ 
semileptonic form factor $f^{D \rightarrow \pi}_+(q^2)$ at $q^2 = 0$,
 with errors a factor of 
two better than in the past.  This combined with recent precision 
measurement of the branching fraction for this process by CLEO-c 
 has allowed for an accurate determination 
of the CKM matrix element $|V_{cd}|$.  Direct determination of 
$|V_{cd}|$ from $D$ semileptonic 
decays is becoming competitive with that from neutrino scattering.
The fact that these two very different processes lead to the same 
$|V_{cd}|$ is a nontrivial consistency check of the Standard Model.

Finally using our values for $|V_{cd}|$ and  $|V_{cs}|$ \cite{dtok}
 plus the most recent $|V_{cb}|_{excl.} 
= 0.0397(10)$ 
from the Fermilab Lattice \& MILC collaborations \cite{vcb}, the most up-to-date 
test of second row unitary from Lattice QCD becomes,
\be
|V_{cd}|^2 +|V_{cs}|^2 + |V_{cb}|^2 = 0.976(50).
\ee
This improves on the PDG2010 value $1.101(74)$ \cite{pdg2010}.

In the future 
we will be reducing the largest errors in Table III by increasing 
statistics and simulating on finer lattices \cite{jona}.  Calculations of 
the full $q^2$ dependence of $f_+^{D \rightarrow K}(q^2)$ and $f_+^{D \rightarrow \pi}
(q^2)$ are also already underway \cite{jona}.
 Furthermore we are 
 working on updating HPQCD's result for the $D$ meson decay constant 
$f_D$ \cite{follana} and on carrying out an independent extraction of $|V_{cd}|$ from 
$D$ leptonic decays \cite{pacs}. 

\vspace{.1in}
{\bf Acknowledgments}: \\
This work was supported by the DOE and NSF in the US, by STFC in the UK and 
by MICINN and DGIID-DGA in Spain. We thank MILC for making their Lattices available. 
Simulations were carried out on facilities of the USQCD collaboration funded by the 
Office of Science of the DOE and at the Ohio Supercomputer Center.

\appendix

\section{Priors and Prior Widths for Two- and Three-Point Correlators}
In this appendix we give sample priors and prior widths used in the fits 
of section III (the reader is referred to reference \cite{bayse} for 
definitions of these terms). 
We turn to the Particle Data Group listings for
guidance in picking priors for energies and energy splittings.  All energies
in Table IV are given in lattices units and are appropriate for ensemble C2.
Numbers for other ensembles can be obtained by rescaling with corresponding
lattice spacings. Prior widths for amplitudes are fixed based on exploratory
initial fits.

\begin{table}[b]
\caption{
Sample priors and prior widths for two- and three-point 
correlator fits. Energies are in lattice units, and this example corresponds to priors used for ensemble C2. 
The notation is the same as in equations (\ref{twopntfit}) and 
(\ref{thrpntfit}).
}
\begin{center}
\begin{tabular}{|c|c|c|}
\hline
  &  prior  &  prior width \\
\hline
\hline
$E^D_0$   &  1.16  & 0.58  \\
$E^D_{j>0} - E^D_{j-1}$  &  0.36  & 0.36 \\
\hline
$E^{\prime D}_0$  &  1.52 & 1.52 \\
$E^{\prime D}_{k>0} - E^{\prime D}_{k-1}$ & 0.36 & 0.36 \\
\hline
$\sqrt{b^D_j}$  & 0.01  & 0.5  \\
\hline
$\sqrt{d^D_k}$  &  0.01  &  0.5  \\
\hline
\hline
$E^\pi_0(000)$   & 0.21  & 0.11  \\
$E^\pi_0(100)$   & 0.38  & 0.19  \\
$E^\pi_0(110)$   & 0.49  & 0.25  \\
$E^\pi_0(111)$   & 0.60  & 0.30  \\
$E^\pi_1(all\;\; mom) - E^\pi_0(all \;\; mom)$  &  0.61  & 0.31 \\
$E^\pi_{j>1}(all \;\; mom) - E^\pi_{j-1}(all \;\; mom)$  &  0.36  & 0.36 \\
\hline
$E^{\prime \pi}_0(100)$  & 0.74 & 0.74 \\
$E^{\prime \pi}_0(110)$  & 0.85 & 0.85 \\
$E^{\prime \pi}_0(111)$  &  0.96 & 0.96 \\
$E^{\prime \pi}_{k>0}(mom > 0) - E^{\prime \pi}_{k-1}(mom > 0)$
 & 0.36 & 0.36 \\
\hline
$\sqrt{b^\pi_j}(all \;\; mom)$  & 0.01  & 0.5  \\
\hline
$\sqrt{d^\pi_k}(mom > 0)$  &  0.01  &  0.5  \\
\hline
\hline
$A_{jk}$, $B_{jk}$, $C_{jk}$, $D_{jk}$ &  0.01  &  0.1 \\
\hline
\end{tabular}
\end{center}
\end{table}



\begin{thebibliography}{99}
\bibitem{fnote1}
For very complicated multi-exponential fits it can happen that 
fit results include fake low lying levels (with amplitudes that are 
consistant with zero) 
 that distort ground state energies and 
 fitting errors.   Care is required to watch out for these spurious levels and 
make sure that fits that exhibit such levels are not trusted.
The ``sequential fitting'' method adopted in this article has been very effective in 
preventing spurious levels from arising.
\bibitem{fnote2}
In reference \cite{dtok} the ``sequential fitting'' method was not employed. 
Instead the number of exponentials in the three-point correlator ansatz was 
kept low, at 1 or 2, and only $N{exp}$ in the two-point correlators increased 
as in Figs. 1 \& 2. 
Good $\chi^2/dof$ were obtained in the simultaneous 
three- and two-point correlator fits and consistant results for 
different $T$ values.  We took this as evidence that excited state contributions 
were suppressed in the three-point correlators compared to in 
two-point correlators.  By simplifying the fit ansatz in this manner,   
we were able to avoid spurious levels from appearing in \cite{dtok}
\end{thebibliography}


\end{document}